# Enhanced stability of the focus obtained by wavefront optimization in dynamical scattering media

**Baptiste Blochet**[1,2,*], **Kelly Joaquina**[1], **Lisa Blum**[1], **Laurent Bourdieu**[2,†], **Sylvain Gigan**[1,†]

[1] *Laboratoire Kastler Brossel, UPMC-Sorbonne Universités, ENS-PSL Research University, CNRS, Collège de France ; 24 rue Lhomond, F-75005 Paris, France*
[2] *IBENS, Département de Biologie, Ecole Normale Supérieure, CNRS, Inserm, PSL Research University, F-75005 Paris, France*
[†] *Senior author*

*baptiste.blochet@lkb.ens.fr*

**Abstract:** Focusing scattered light using wavefront shaping provides interesting perspectives to image deep in opaque samples, as e.g. in nonlinear fluorescence microscopy. Applying these technics to in vivo imaging remains challenging due to the short decorrelation time of the speckle in depth, as focusing and imaging has to be achieved within the order of the decorrelation time. In this paper, we experimentally study the focus lifetime after focusing through dynamical scattering media, when iterative wavefront optimization and speckle decorrelation occur over the same timescale. We show experimental situations with heterogeneous stability of the scattering sequences, where the focus presents significantly higher stability than the surrounding speckle.

## 1. Introduction

In recent years, several wavefront shaping techniques were developed to partially compensate for the scattering induced by a disordered medium and to form a diffraction limited focus using the scattered light [**1**], possibly at depth non-invasively [**2**]. However, a major limitation to the application of these techniques to the imaging of real biological systems is the temporal decorrelation induced by minute changes of the optical index inhomogeneity: the decorrelation time of biological tissues can be in the millisecond range [**3-4**]. As a consequence, fast wavefront shaping systems are required for focusing light in these systems [**4-5**] and the lifetime of the formed focus is also limited by this decorrelation time [**6**]. Two main approaches have been developed to focus within this decorrelation time. On the one hand, digital optical phase conjugation (DOPC) relying on the phase conjugation of a measured wavefront is a fast non-iterative technique capable of focusing in the millisecond range [**7-8**]. On the other hand, iterative optimizations can focus almost as fast through a scattering medium [**9-12**]. While DOPC methods are very appealing to tackle fast decorrelating media, optimization methods remain inescapable in many bio-imaging scenarios, particular based on two-photon fluorescence [**13-16**].

So far, the lifetime of the focus obtained after wavefront shaping has only been studied for DOPC [**6-7**]. In particular, it has been shown experimentally and theoretically, that the temporal correlation function in intensity $g_2(t)$, used to quantify the decorrelation dynamics of the speckle, also gives the temporal decay of the focus intensity after phase conjugation. An analogous experiment has also been conducted with iterative optimization in the frequency domain [**17**], but in a stationary medium where the decorrelation is induced by tuning the incident wavelength rather than by a physical displacement of the scatterers. In this paper, the authors demonstrated experimentally and theoretically that the focus intensity degradation is proportional to the spectral correlation function in intensity of the speckle after shifting the laser frequency. In all these works [**6, 7, 17**], the medium can be considered as static during the wavefront shaping procedure resulting in a proportionality between speckle decorrelation and focus degradation. At the opposite, if the wavefront correction procedure is much slower than the decorrelation time, focusing of the scattered photons cannot be achieved [**9**].

However, an interesting question remains: what happens if we perform an iterative optimization through a dynamical scattering media, where wavefront correction and decorrelation occur over similar timescales? In this paper, we investigate which scattering medium properties (mean decorrelation time, width of the decorrelation time distribution) impact the characteristic lifetime of the focus obtained in such a scenario. In particular, we show that there are experimental situations where the focus can be significantly more stable than the surrounding speckle pattern, a situation that arises for instance when the medium has heterogeneous stability. Finally, we will experimentally demonstrate that this phenomenon can be observed in acute brain slices.

## 2. Principle

The output speckle pattern formed after a scattering media can be seen as a coherent sum of scattering sequences (i.e. the diffusion path of a photon in the medium) with random phases and amplitudes. Inside a dynamical scattering medium, each scattering sequence presents also a given time stability. The temporal intensity correlation function $g_2(t)$ characterizes the temporal decorrelation of these sequences. For example, the slope at the origin of this function gives the mean decorrelation time, which is the mean time over which the scattering sequences change.

Performing wavefront shaping with a spatial light modulator (SLM) means adding the appropriate phase to the incident wavefront to sum constructively some sequences at a desire target to form a sharp focus. If the sequences selected to focus through a dynamical scattering medium present the same temporal stability as the full set of sequences forming the speckle, the focus and the speckle should have the same decorrelation dynamics, as it was observed in DOPC. Hypothetically, if more stable scattering sequences could be favored during wavefront shaping (by selecting either more stable scatterers or shorter sequences), the focus should present a higher stability than the initial speckle before optimization.

For a continuous iterative optimization, if decorrelation and optimization occur over similar timescales, it is not clear which scattering sequences will be used to form a focus. We experimentally investigated this question with a custom fast wavefront shaping system [**9**]. We used this setup to focus through synthetic homogeneous scattering media of various stabilities and scattering strengths, but also stratified media with a "static" and a "dynamic" part, as it can be the case for biological samples. In this latter case, the width of the temporal stability distribution of the different scattering "sequences" in the medium can be broad and heterogeneous. We studied the stability of the focus in these various scenarios and investigated under which conditions the focus could present an enhanced stability.

Fig. 1 describes the experimental wavefront shaping setup. A phase only spatial light modulator (Kilo-DM segmented, Boston Micromachines) shapes the incident wavefront of a CW laser λ=532 nm (Coherent Sapphire). The SLM is conjugated to the back focal plane of a microscope objective (10x, 0.25), which illuminates a scattering sample. The polarized output speckle is simultaneously imaged onto a CCD camera (Allied Vision Technologies Manta G-046B) and on a mono-detector (PMT, Hamamatsu H10721-20). A continuous iterative wavefront optimization algorithm is implemented to maximize the intensity of one speckle grain collected by the PMT. In short, the optimization is obtained using the Hadamard input basis. At each iteration, half of the pixels are modulated in phase, while the PMT signal is monitored, and the optimal phase is added to the correction mask. We combined a fast acquisition card (NI PXIe-6361) and a fast FPGA board (NI PXIe-7962R) to reach a speed of 4,1 kHz per mode [**9**]. For all of experiments, the full Hadamard basis is successively optimized 5 times in 1.25 s.

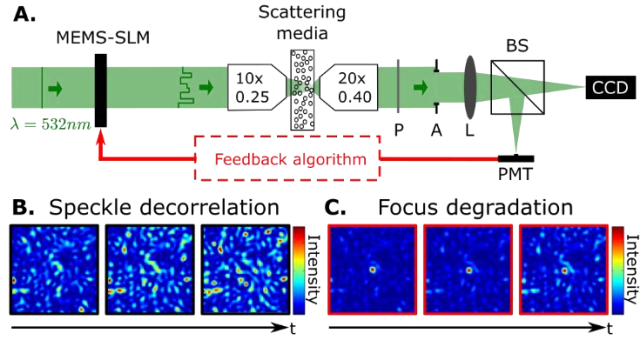

Fig. 1. **(A)** Experimental setup. P: polarizer; A: aperture; L: lens (focal length = 150mm); BS: beamsplitter; MEMS-SLM; MEMS-based spatial light modulator. The wavefront of a collimated laser beam (532 nm) is modulated by a phase-only spatial light modulator. The phase mask is imaged on the back aperture of a microscope objective and focused into a scattering sample. A second microscope objective images the output speckle using a beamsplitter on a CCD camera and on a PMT. The PMT collects the intensity of one speckle grain through an optical fiber. An iris controls the aperture size to match the speckle grain size with the diameter of the fiber. A polarizer selects one polarization state of the output speckle. The PMT signal is acquired by a DAQ board and sent to a FPGA board. During the optimization algorithm, the FPGA board computes the optimal phase for a given Hadamard mode, adds it to the current phase mask of the SLM and applies the new mask to the SLM. One mode optimization takes 243 µs. **(B)** A stack of speckle patterns is recorded over time to characterize the decorrelation dynamics of the speckle. **(C)** After ending the optimization, the focus degrades in time due to the speckle decorrelation.

Our experimental system is capable of measuring successively the temporal correlation function in intensity $g_2(t)$ and the focus degradation after ending the optimization. Indeed, the PMT signal is simultaneously collected onto a computer to measure the focus degradation after ending the optimization. The temporal correlation function in intensity is measured with the CCD camera [18]. To compare the normalized focus degradation and the $g_2$ function, we fit them with an exponential function [19]:

$$(1-\Gamma) * \exp(-t/\tau) + \Gamma \quad (1),$$

where $\tau$ is the mean decorrelation time of the focus / speckle and $\Gamma$ quantifies the proportion of static sequences that can contribute to the focus / speckle. Therefore, we can compare for all experiments the mean decorrelation time (for the speckle and the focus) of the dynamic scattering sequences using $\tau$ and, if they exist, the ratio of static scattering sequences using $\Gamma$.

## 3. Results

We have used different samples in order to understand in which situation a stable focus can be formed. We studied first the case of a colloidal solution in multiple scattering regime, where difference of sequence time stability results from difference in their length. Due to experimental constrains, the width of the path length distribution of such a media (and therefore its temporal stability distribution) is not tunable. To study the impact of the width of the temporal stability distribution, we then designed a second category of samples, composed of a thin dynamical layer above a static layer in multiple scattering regime. In this medium, part of the light travels ballistically through the dynamical layer. Therefore, static scattering sequences exist through the sample. By varying the size of the scatterers inside the colloidal solution, the time stability of the dynamical scattering sequences could be tuned. For small polystyrene beads, light scattered was highly unstable. In this situation our optimization scheme was only capable of compensating for the static sequences. On the other hand, for large polystyrene beads, the decorrelation was slower. In this last situation, our system was both capable of compensating static and dynamical scattering. Finally, we achieved the same experiment through acute brain slices from the brainstem.

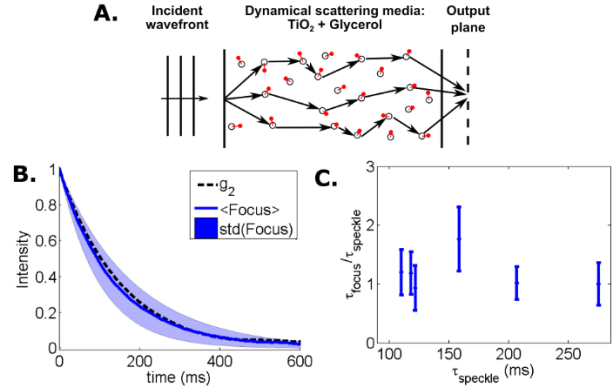

Fig 2. Focus stability through a monodisperse colloidal solution. **(A)** Scheme of the scattering process. When propagating inside a scattering medium, light will travel through many scattering sequences and will then interfere to form a speckle pattern. For a dynamical medium, sequences will change in time leading to the decorrelation of the speckle. **(B)** Evolution in time of the focus after ending the optimization through a medium with a mean decorrelation time of 140 ms: average (solid line) and standard deviation (blue region) over 500 realizations. Dotted line: intensity correlation function ($g_2$) of the speckle. Averaged over a large number of realizations, the focus presents the same stability as the speckle, even if each individual realization doesn't. **(C)** Ratio of the focus mean decorrelation time and of the speckle decorrelation time for different speckle decorrelation times. In average, the focus presents the same stability as the speckle through a monodisperse colloidal solution.

### 3.1 Monodisperse colloidal solution

The first sample used was a 500 μm thick solution of TiO2 (Sigma Aldrich 224227) in glycerol with a mass concentration of 20 g/l ($l_s$ = 70 μm and $l^*$ = 200 μm) [9]. Light propagation through the sample is therefore in a regime of multiple scattering. A schematic of a few dynamical sequences is illustrated in fig 2.a. The decorrelation time of a scattering sequence in a monodisperse solution is directly related to the number of scattering events [20]. Furthermore, tuning the temperature modifies the viscosity of the sample, thus allowing to tune its mean decorrelation time.

The temperature of the sample was first adjusted at 16 ° C to obtain an average decorrelation time of the speckle of $\tau_{speckle}$ = 140 ms. The resulting mean focus degradation and its standard deviation are shown on fig 2.b. For this dynamical sample, a mean focus decorrelation time of $\tau_{focus}$ = 140 ms with a standard deviation of 43 ms was measured over 500 realizations.

We then measured the average value of the focus decorrelation time and its standard deviation for different medium stabilities ranging from 100 ms to 300 ms (obtained by changing the viscosity of the solution via the sample temperature). In fig. 2.c, the ratio $\tau_{focus}$ / $\tau_{speckle}$ is shown. This ratio is constant and close to unity for all tested stabilities. Interestingly, an individual realization of the focus may have a characteristic time different from the one of the speckle, but on average they are identical.

So far, using monodisperse colloidal solutions, we didn't find any optimization procedures that may favor stable scattering sequences. The weak difference between time stability of the difference scattering sequences might explain this observation. To investigate further, we synthetized dynamical scattering media with a wider time stability distribution of the different scattering "sequences".

### 3.2 Combination of layers of static and dynamic scatterers

In a second experiment, we synthetized dynamical scattering media that exhibits a wider range of sequence stability (see Figure 3.a for a diagram of the scattering sequences existing in our media). By superimposing two scattering media, a thick static layer and a thin dynamical layer, we were able to control the percentage of dynamical scattering sequences exiting the sample.

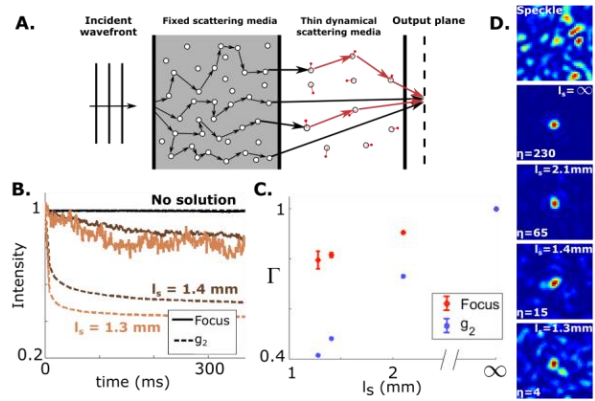

**FIG 3.** Focusing through two layers of scattering media (a static scattering medium and a fast dynamical scattering medium) **(A)** Scheme of the scattering sample. Light is first multiply scattered by a fixed scattering layer (Teflon, in grey). Then light encounters only few scattering events by propagating through a dynamical scattering layer (aqueous colloidal solution of polystyrene: Polybead® Carboxylate Microspheres 0.35 μm). Part of the light propagates ballistically through this dynamical layer (black arrows); the other sequences (in red) decorrelate due to the motion of the scatterers. **(B)** Comparison between focus degradation (solid lines) and speckle decorrelation (dotted lines) for different scattering mean free paths of the colloidal solution. The mean speckle decorrelation time, in presence of the colloidal solution, is below 1 ms. The optimization process isn't fast enough to compensate for this dynamical scattering. Therefore, the sequences contributing to the focus are mostly static sequences. **(C)** Mean value of the plateau $\Gamma$ after decorrelation for the focus (red diamonds) and the speckle (blue square) for different scattering mean free path of the colloidal solution. Error bars represent the 95% confidence bounds of the fit. **(D)** Top: speckle pattern measured before optimization and then respectively from top to bottom: CCD images (320*320 μm$^2$) acquired after a stable focus is obtained for different scattering samples ($l_s = \infty$, 2.1 mm, 1.4 mm, 1.2 mm). As the scattering strength of the dynamical layer increases, the enhancement decreases. In this situation, the dynamical scattering layer can be seen as an extra experimental noise which reduces the final enhancement.

In this experiment, we wanted to investigate which sequences (fixed or dynamical) will form a focus after optimization. In a first experimental situation, we designed a sample where the dynamical sequences were decorrelating too fast to be corrected by our wavefront shaping system. In a second situation, we designed a sample where the dynamical sequences were slower and may be compensated by our system [**9**].

The first medium was a solution of polystyrene beads (Polybead® Carboxylate Microspheres 0.35 μm) in water positioned over a thick static scattering medium. The thickness of the layer of the dynamical scattering solution was 1 mm. The percentage of ballistic photons through the scattering solution was controlled by adjusting the polystyrene beads concentration. Using Mie theory, the concentration required to obtain a given mean free diffusion path in the dynamical medium can be computed. This percentage ranges from 46% ($l_s$ = 1.3 mm) to 100% (no scattering solution). The addition of a strongly scattering static layer ensures that we are overall in the multiply scattering regime.

For each solution, the measurement of $g_2$ confirms that the speckle resulting from dynamically scattered photons decorrelates in less than a millisecond (see Figure 3.b, dashed lines). We also observe that the $g_2$ function reaches a plateau ($\Gamma_{speckle}$) for large decorrelation times, indicating that static sequences contribute to the speckle pattern. For each sample, the decorrelation of the focus averaged over 100 realizations is plotted in Figure 3.b (solid line). Yet, for all tested concentrations of colloidal solutions (figure 3.b), the wavefront correction system was able to form a focus, which was decorrelating slower than the speckle and was eventually reaching a plateau. As our system is not fast enough to compensate for the dynamically scattered photons, the value of the plateau ($\Gamma_{focus}$) was larger than the one measured for the speckle ($\Gamma_{speckle}$), showing that the focus contains a larger amount of static sequences

(figure 3.b). As the scattering mean free path decreases, more and more photons were scattered by the dynamical layer (figure 3.c). Nevertheless, in all cases, the focus obtained by wavefront shaping was mostly formed by static scattering sequences. Interestingly, some slowly decorrelating scattering sequences are also contributing to the focus. These more stable sequences are probably snake-like sequences that encounter only very few forward scattering events. Finally, we have seen that, through these samples, the mode distribution used to form a focus is very different from the one of the speckle. Moreover, figure 3.d shows that the enhancement of the focus intensity by optimization was larger for smaller bead concentrations in the dynamical samples. Indeed, the dynamical scattering speckle can be seen as an extra noise that reduces the enhancement [**21**].

We then synthetized a similar dynamical medium but with larger colloidal beads (Polybead® Carboxylate Microspheres 1 µm). The larger beads sustained higher viscosity forces, which increased the temporal stability of the dynamical scattering sequences. By tuning the concentration (therefore $l_s$), we simultaneously controlled the percentage of fixed sequences that exited from the sample and the mean decorrelation time of the dynamical sequences. For all prepared samples, the mean speckle decorrelation times ranged from 100 ms to 500 ms and the proportion of fixed scattering sequences ranged from almost 0 to 80%. Our wavefront shaping system should therefore be capable of optimizing the phase of the wavefront travelling through any of these sequences [**9**].

The results are presented in figure 4 for $l_s$ ranging from 0.6 mm to 2.3 mm. The blue squares and the red diamonds indicate the average proportion of static sequences (fig 4.a) and the mean decorrelation time (fig 4.b) respectively for the speckle and the focus. The data obtained for the focus were averaged over 100 realizations. For large $l_s$, most of the sequences contributing to the focus or the speckle are static and therefore the focus is as stable as the speckle. As $l_s$ decreases, the sample exhibits larger differences in term of sequence stability and the wavefront shaping process tends to favor static scattering events and the more stable sequences among the dynamical ones: the focus contains more static sequences ($\Gamma_{focus}$) than the speckle ($\Gamma_{speckle}$) and its lifetime ($\tau_{focus}$) is larger than the one of the speckle ($\tau_{speckle}$) by a factor 2 in the best case ($l_s$ = 0.6 mm).

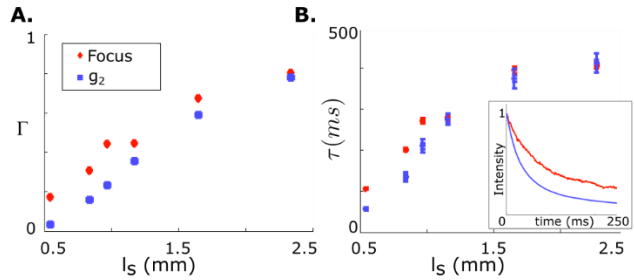

Fig 4. Focusing through two layers of scattering media (a static medium and a slow dynamical scattering medium). The scheme of the experiment is similar to the one shown on Fig. 3. The dynamical scattering medium is an aqueous colloidal solution of polystyrene beads (Polybead® Microspheres 1.00 µm). Due to larger polystyrene beads, the mean scattering decorrelation time is slower ranging from 55 ms to 405 ms, as compared to the case shown in Fig. 3. Here, the optimization procedure is fast enough to compensate for the dynamical scattering. **(A)** Mean position of the plateau $\Gamma$ after decorrelation for the speckle (blue squares) and for the focus (red diamonds), for different scattering mean free path of the colloidal solution. Error bars represent the 95% confidence bounds of the fit. In average, the focus generated after wavefront shaping is formed by more static scattering sequences than the surrounding speckle. **(B)** Mean lifetime of the focus (red diamonds) and mean decorrelation time of the speckle (blue squares) for different scattering mean free paths of the colloidal solution. Error bars represent the 95% confidence bounds of the fit. In average, for small $l_s$, the dynamical sequences interfering constructively at the focus show a larger stability than the surrounding speckle. <u>Inset</u>: Example of a focus degradation (red line) and speckle decorrelation (blue line) for a colloidal solution with $l_s$ = 0.6 mm.

These more stable sequences are probably snake-like sequences that encounter only few forward scattering events. The enhancement in intensity follows a linear trend similar to the one previously reported [**9**] ranging from 20 for $\tau_{speckle}$ =50 ms to 120 for $\tau_{speckle}$ =450 ms.

To conclude, the key element to form a focus with stable sequences seems to be the width of the stability time distribution of the different scattering sequences. The broader the time distribution of the different scattering sequences, the more stable sequences the focus contains and the larger its lifetime.

### 3.3 Biological samples

As a last experiment, we investigated whether our optimization algorithm allows achieving in biological tissues a focus more stable than the speckle. This would of course be very beneficial to perform non-linear imaging after wavefront correction, since an increased focus stability would provide additional time for the formation of a fluorescence image.

Our sample was a 300 µm thick acute slice of mouse brain ($l_s \sim 40$ µm [**22**]). To keep the slices alive, a stream of a solution of 125 mM NaCl, 2.5 mM KCl, 2 mM $CaCl_2$, 1 mM $MgCl_2$, 1.25 mM $NaH_2PO_4$, 26 mM $NaHCO_3$ and 25 mM glucose, bubbled with 95% $O_2$ and 5% $CO_2$, was imposed around the wafer [**23**]. Every effort has been made to keep the brain slices alive for the duration of the experiment. The scheme of the system used to maintain the slice acute is shown on fig. 5.a and a typical widefield image of a slice is shown on fig. 5.b.

Figure 5.c shows (solid blue line) the temporal correlation function of the speckle. A first rapid decorrelation (slope at the origin of ~205 ms) is followed at long times by a slower decorrelation with a typical timescale of the order of 10 s. We did not study here the microscopic origin of these different decorrelation times. As in the previous experiment, we observed that the focus obtained by optimization (solid red line) is on average (over 500 realizations) more stable than the speckle and presents an average enhancement of $33 \pm 10$. The red dotted line and the red crossed line show respectively the least stable focus and the most stable focus. These two optimizations led to identical enhancement of the order of 30. It seems that there are no benefits (or drawbacks) in term of enhancement to favor stable sequences. Interestingly, the most stable focus generated here seems to be almost perfectly stable (over tens of seconds).

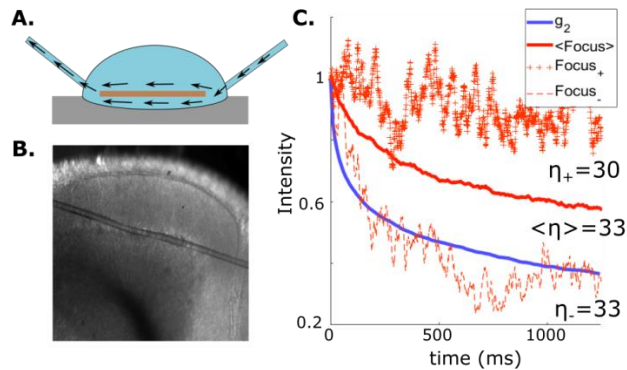

Fig 5. (**A**) Scheme of the setup to maintain acute brain slices. A slice is immersed in an oxygenized buffer which is renewed with a flux. (**B**) Oblique wide field image (4.6*4.6 $mm^2$) of a typical acute brain slice (cerebellum). (**C**) Focusing through an acute mouse brain slice (brainstem) of 300 µm. Intensity correlation of the speckle: blue; average focus degradation (red, <Focus>), more stable degradation ($Focus_+$) and less stable degradation ($Focus_-$). The scattering sequences show a fast decorrelation (~100 ms) followed by a slow one (~10 s). In average the optimization process promotes the most stable sequences at the focus. In the best case, the focus is only formed with stable sequences. On the contrary, in the worst case, the focus degradation follows the speckle decorrelation.

## 4. Discussion and conclusion

We have shown that, in contrast with previous wavefront shaping experiments, the focus does not always have the same stability as the surrounding speckle. We have shown in particular that in specific conditions an increase of the focus stability by a factor 2 (Fig. 4) up to several orders of magnitude (Fig. 3) is obtained, as compared to the speckle stability.

Our interpretation of this results is the following. At each iteration of the optimization, some stable and dynamic sequences are corrected to interfere constructively to the focus. Rapidly, the dynamic sequences decorrelate and do not contribute to the focus anymore, while the stable sequences still do. Therefore, iteration after iteration, more and more stable sequences accumulate leading to an enhanced stability of the focus. Ultimately, all the SLM mode available could compensate only for the stable sequences (as observed in Fig. 5.c).

The key to achieve a more stable focus seems to be the width of the time stability distribution of the scattering sequences. The wider this distribution, the easier it is to promote stable diffusion sequences by optimization. On the contrary, a narrow distribution (as through a monodisperse solution) did not allow, for an iterative optimization of the wavefront, the selection of more stable sequences, at least for the range of parameters investigated.

Another key element to obtain a more stable focus is the speed at which the optimization is done. If the optimization is too fast compared to the decorrelation time of the medium, this effect does not appear (as in DOPC). In our case, we observed that if the optimization time is of the order of the stability times of the medium, a focus more stable than the speckle can be formed.

The influence of these two parameters (width of the time stability distribution of the scattering sequences, speed at which the optimization is done) remains difficult to analyze experimentally and further numerical studies (beyond the scope of this paper) may be required to fully describe their respective role. Additional studies of the impact of the optimization algorithm on the stability of the focus could highlight optimization strategies that can further promote the emergence of a more stable focus.

We believe these results are of great interest particularly for biomedical imaging. For instance, during in vivo imaging of a mouse brain (the skull having been removed and replaced by a glass coverslip), part of the light propagates through or around blood vessels, thus imposing a very rapid decorrelation of the speckle. Despite this, a wavefront correction system should be able to focus the photons scattered by static structures or having a slow dynamic (cells, myelinated axons, …), if the fraction of dynamically scattered light remains low.

Another important scenario is imaging through the skull. The skull would then act as a nearly static scatterer and the brain tissue would be the dynamical scatterer. In this case, we expect the wavefront correction system to preferentially correct the scattering by the skull.

An interesting case is the correction of the wavefront in the presence of a ballistic but aberrant wavefront inside a tissue. The scattered light rapidly decorrelates whereas the aberrated light will be relatively stable. A correction of the wavefront would then preferentially correct aberrations.

One last perspective could be to extend this study to the broadband regime. Mounaix et al demonstrated a selection of short scattering sequences by exploiting the short coherence length of a pulsed laser through a homogeneous scattering media [**24**]. By exploiting this effect, one could obtain a further increase in the stability of the focus.

**Funding**. H2020 European Research Council (ERC) (278025, 724473);

Université Pierre et Marie Curie (UPMC) ; Agence Nationale de la Recherche (ANR-15-CE19-0011-01 and ANR-15-CE19-0011-03  ALPINS, ANR-10-LABX-54 MEMOLIFE, ANR-10-IDEX-0001-02 PSL* Research University and ANR-10-INSB-04-01 France-BioImaging infrastructure.

**Acknowledgment**. We thank Claudio Moretti, Dimitri Dumontier, Stephane Dieudonné and Mariano Casado for valuable help with the acute brain slices experiment and Cathie Ventalon for advices and comments on the project. S. G. is a member of the Institut Universitaire de France. B. B. was funded by a Ph.D. fellowship from UPMC under the program "Interface pour le Vivant".

**References**

1.  Rotter, S., & Gigan, S. (2017). Light fields in complex media: Mesoscopic scattering meets wave control. Reviews of Modern Physics, 89(1), 015005.
2.  Horstmeyer, R., Ruan, H., & Yang, C. (2015). Guidestar-assisted wavefront-shaping methods for focusing light into biological tissue. Nature photonics, 9(9), 563.
3.  Qureshi, M. M., Brake, J., Jeon, H. J., Ruan, H., Liu, Y., Safi, A. M., ... & Chung, E. (2017). In vivo study of optical speckle decorrelation time across depths in the mouse brain. Biomedical optics express, 8(11), 4855-4864.
4.  Liu, Yan, et al. "Optical focusing deep inside dynamic scattering media with near-infrared time-reversed ultrasonically encoded (TRUE) light." Nature communications6 (2015): 5904.
5.  Stockbridge, C., Lu, Y., Moore, J., Hoffman, S., Paxman, R., Toussaint, K., & Bifano, T. (2012). Focusing through dynamic scattering media. Optics express, 20(14), 15086-15092.
6.  Jang, M., Ruan, H., Vellekoop, I. M., Judkewitz, B., Chung, E., & Yang, C. (2015). Relation between speckle decorrelation and optical phase conjugation (OPC)-based turbidity suppression through dynamic scattering media: a study on in vivo mouse skin. Biomedical optics express, 6(1), 72-85.
7.  Liu, Y., Ma, C., Shen, Y., Shi, J., & Wang, L. V. (2017). Focusing light inside dynamic scattering media with millisecond digital optical phase conjugation. Optica, 4(2), 280-288.
8.  Wang, D., Zhou, E. H., Brake, J., Ruan, H., Jang, M., & Yang, C. (2015). Focusing through dynamic tissue with millisecond digital optical phase conjugation. Optica, 2(8), 728-735.
9.  Blochet, B., Bourdieu, L., & Gigan, S. (2017). Focusing light through dynamical samples using fast continuous wavefront optimization. Optics letters, 42(23), 4994-4997.
10. Cui, M. (2011). A high speed wavefront determination method based on spatial frequency modulations for focusing light through random scattering media. Optics express, 19(4), 2989-2995.
11. Feldkhun, D., Tzang, O., Wagner, K. H., & Piestun, R. (2019). Focusing and scanning through scattering media in microseconds. Optica, 6(1), 72-75.
12. Tzang, O., Niv, E., Singh, S., Labouesse, S., Myatt, G., & Piestun, R. (2018). Wavefront shaping in complex media at 350 KHz with a 1D-to-2D transform. arXiv preprint arXiv:1808.09025.
13. Tang, J., Germain, R. N., & Cui, M. (2012). Superpenetration optical microscopy by iterative multiphoton adaptive compensation technique. Proceedings of the National Academy of Sciences, 109(22), 8434-8439.
14. Papadopoulos, I. N., Jouhanneau, J. S., Poulet, J. F., & Judkewitz, B. (2017). Scattering compensation by focus scanning holographic aberration probing (F-SHARP). Nature Photonics, 11(2), 116.
15. Galwaduge, P. T., Kim, S. H., Grosberg, L. E., & Hillman, E. M. C. (2015). Simple wavefront correction framework for two-photon microscopy of in-vivo brain. Biomedical optics express, 6(8), 2997-3013.
16. Ji, N., Milkie, D. E., & Betzig, E. (2010). Adaptive optics via pupil segmentation for high-resolution imaging in biological tissues. Nature methods, 7(2), 141.
17. Van Beijnum, F., Van Putten, E. G., Lagendijk, A., & Mosk, A. P. (2011). Frequency bandwidth of light focused through turbid media. Optics letters, 36(3), 373-375.
18. Wong, A. P., & Wiltzius, P. (1993). Dynamic light scattering with a CCD camera. Review of Scientific Instruments, 64(9), 2547-2549.
19. Brake, J., Jang, M., & Yang, C. (2016). Analyzing the relationship between decorrelation time and tissue thickness in acute rat brain slices using multispeckle diffusing wave spectroscopy. JOSA A, 33(2), 270-275.
20. Maret, G., & Wolf, P. E. (1987). Multiple light scattering from disordered media. The effect of Brownian motion of scatterers. Zeitschrift für Physik B Condensed Matter, 65(4), 409-413.

21. Vellekoop, I. M., & Mosk, A. P. (2008). Phase control algorithms for focusing light through turbid media. Optics communications, 281(11), 3071-3080.
22. Jacques, S. L. (2013). Optical properties of biological tissues: a review. Physics in Medicine & Biology, 58(11), R37.
23. Paoletti, P., Ascher, P., & Neyton, J. (1997). High-affinity zinc inhibition of NMDA NR1–NR2A receptors. Journal of Neuroscience, 17(15), 5711-5725.
24. Mounaix, M., de Aguiar, H. B., & Gigan, S. (2017). Temporal recompression through a scattering medium via a broadband transmission matrix. Optica, 4(10), 1289-1292.